\documentclass[final]{ias2}

\usepackage{graphicx} 
\usepackage{multirow}
\usepackage{array}
\usepackage{natbib}
\usepackage{bm}

\usepackage{hyperref} 

\newcommand{\EQ}{\begin{equation}}
\newcommand{\EN}{\end{equation}}
\newcommand{\EQA}{\begin{eqnarray}}
\newcommand{\ENA}{\end{eqnarray}}
\newcommand{\eq}[1]{(\ref{#1})}

\newcommand{\Fig}[1]{Figure~\ref{#1}}

\newcommand{\bra}[1]{\langle #1\rangle}

\newcommand{\AAA}{\bm{A}}

\newcommand{\BB}{\bm{B}}

\newcommand{\hh}{\bm{h}}
\newcommand{\kk}{\bm{k}}

\newcommand{\xx}{\bm{x}}

\newcommand{\nab}{\mbox{\boldmath $\nabla$} {}}

\newcommand{\ii}{{\rm i}}

\newcommand{\dd}{{\rm d} {}}

\def\Rm{R_{\rm m}}

\def\kf{k_{\rm f}}

\def\kmean{k_{\rm m}}
\def\urms{u_{\rm rms}}
\def\Beq{B_{\rm eq}}

\def\etat{\eta_{\rm t}}

\newcommand{\uG}{\,\mu{\rm G}}

\newcommand{\yan}[3]{, {\em Astron.\ Nachr.\ }{\bf #2}, #3 (#1)}

\newcommand{\ypnas}[3]{, {\em Proc.\ Nat.\ Acad.\ Sci.}{\bf #2}, #3 (#1)}
\newcommand{\yana}[3]{, {\em Astron.\ Astrophys.\ }{\bf #2}, #3 (#1)}

\newcommand{\ysci}[3]{, {\em Science }{\bf #2}, #3 (#1)}
\newcommand{\ysph}[3]{, {\em Solar Phys.\ }{\bf #2}, #3 (#1)}
\newcommand{\yjetp}[3]{, {\em Sov.\ Phys.\ JETP }{\bf #2}, #3 (#1)}

\newcommand{\ymn}[3]{, {\em Monthly Notices Roy.\ Astron.\ Soc.\ }{\bf #2}, #3 (#1)}

\newcommand{\yjfm}[3]{, {\em J.\ Fluid Mech.\ }{\bf #2}, #3 (#1)}
\newcommand{\ypr}[3]{, {\em Phys.\ Rev.\ }{\bf #2}, #3 (#1)}
\newcommand{\ypre}[3]{, {\em Phys.\ Rev.\ E }{\bf #2}, #3 (#1)}

\newcommand{\yprl}[3]{, {\em Phys.\ Rev.\ Letters }{\bf #2}, #3 (#1)}

\newcommand{\yprs}[3]{, {\em Proc.\ Roy.\ Soc.\ Lond.\ }{\bf #2}, #3 (#1)}

\newcommand{\yptrsa}[3]{, {\em Phil.\ Trans.\ Roy.\ Soc.\ Lond., Ser.\ A }{\bf #2}, #3 (#1)}

\newcommand{\yapj}[3]{, {\em Astrophys.\ J.\ }{\bf #2}, #3 (#1)}

\newcommand{\ypp}[3]{, {\em Phys.\ Plasmas }{\bf #2}, #3 (#1)}

\newcommand{\ypf}[3]{, {\em Phys.\ Fluids }{\bf #2}, #3 (#1)}

\newcommand{\ygafd}[3]{, {\em Geophys.\ Astrophys.\ Fluid Dynam. }{\bf #2}, #3 (#1)}

\newcommand{\yjour}[4]{, {\em #2} {\bf #3}, #4 (#1)}
\newcommand{\yproc}[5]{, in {\em #3} edited by #4, pp.\ #2, #5 (#1)}
\newcommand{\ybook}[3]{ {\em #2}.\ #3 (#1)}

\begin{document}

\markboth{Chandrasekhar--Kendall functions in astrophysical dynamos}
{Axel Brandenburg}

\title{Chandrasekhar--Kendall functions in astrophysical dynamos}

\author[sin,ain]{Axel Brandenburg} 
\email{brandenb@nordita.org}
\address[sin]{NORDITA, AlbaNova University Center,
Roslagstullsbacken 23, SE 10691 Stockholm, Sweden}
\address[ain]{Department of Astronomy, Stockholm University,
SE 10691 Stockholm, Sweden}

\begin{abstract}
Some of the contributions of Chandrasekhar to the field of magnetohydrodynamics
are highlighted.
Particular emphasis is placed on the Chandrasekhar--Kendall functions
that allow a decomposition of a vector field into right- and left-handed
contributions.
Magnetic energy spectra of both contributions are shown for a new set
of helically forced simulations at resolutions higher than what has been
available so far.
For a forcing function with positive helicity, these simulations show a
forward cascade of the right-handed contributions to the magnetic field
and nonlocal inverse transfer for the left-handed contributions.
The speed of inverse transfer is shown to decrease with increasing value
of the magnetic Reynolds number.
\end{abstract}


 
\maketitle


\section{Introduction}

During the 1950s, Chandrasekhar was deeply immersed in studying problems
in hydrodynamics and especially magnetohydrodynamics (MHD), which is the
study of flows of electrically conducting media such as liquid metals and
ionized gases or plasmas.
Between 1950 and 1960, roughly 2/3 of his papers 
were related to hydrodynamic and magnetohydrodynamics stability,
turbulence, and aspects of dynamo theory.
Chandra's scientific activities during that period are reviewed in detail
by Parker \cite{Par96}, who draws a close connection between Chandra's
interests in Heisenberg's theory of turbulence \cite{Hei48} and the then
discovered polarization of starlight in the galaxy \cite{Hal49,Hil49}.
This led to a collaboration between Chandrasekhar and Fermi \cite{CF53},
where they estimated the galactic field strength to be $6$--$7\uG$,
which is close to the commonly accepted value today \cite{FLK08}.
Parker also suggests that observations of magnetic fields in the Sun
\cite{BB55} and other stars \cite{Bab47} may have contributed to Chandra's
interest in magnetohydrodynamic stability and perhaps dynamo theory.

Chandrasekhar's name is not normally connected with dynamo theory.
Indeed, at the time there was still great disbelief in dynamo theory,
given that Cowling \cite{Cow33} and others, including Chandra himself
\cite{Cha56_axisym}, proved dynamos impossible under the assumption of
axisymmetry of the magnetic field.
After all, Chandra's work in that field was concentrated to the period just
before Herzenberg \cite{Her58} produced the first existence proof of
dynamos.
Therefore, Chandrasekhar kept his mind open for alternative explanations.
In 1956 this led him to the question \cite{Cha56_slowdecay} ``Can the
time of decay of 14,000 years, in the absence of internal motions, be
prolonged to 500,000 years (say) by velocities of reasonable magnitude
and patterns?'', which he answered tentatively with ``Yes''.
He discussed this finding in connection with possible solutions to the
problem of the geomagnetic field and even the field observed in sunspots.
Now we know that the physically correct and relevant understanding, at
least for the Sun, came from Parker's work \cite{Par55} of 1955.
However the special significance of Parker's early work was perhaps not
yet fully appreciated when Chandrasekhar \cite{Cha56_slowdecay} in 1956
merely noted that ``the possibility, nevertheless, of dynamo action
has been explored, intensively, in recent years by Bullard \& Gellman
\cite{BG54}, Parker \cite{Par55}, and by Elsasser \cite{Els55,Els56}.''

Chandrasekhar's most cited publication is his book on
{\it Hydrodynamic and hydromagnetic stability} \cite{Cha61}
which, in 2010, received about 140 citations on ADS.
The citation rate has been increasing and was about
70 citations some 10 years earlier.
He studied the stability of virtually any imaginable combination of
systems in slab and spherical geometries, with and without rotation,
with and without magnetic fields, as well as systems with self-gravity.
In astrophysics, hydrodynamic and hydromagnetic instabilities are
important sources of turbulence.
Unlike atmospheric and wind tunnel turbulence, which is typically driven
through boundary layer instabilities, turbulence in astrophysical settings
tends to be driven by instabilities within the volume.
Important examples include Rayleigh-B\'enard convection as well as the
magneto-rotational instability.
Chandrasekhar has contributed to both types of instabilities.
It is now well accepted that at large enough Reynolds numbers, the resulting
flows tend to become turbulent and are invariably subject to yet another
instability: the dynamo instability.
Of particular significance is the case of large-scale dynamo action.
All known mechanisms produce helical large-scale magnetic fields that tend
to be nearly force-free.
This leads to yet another of Chandra's favorite topics at the time,
namely the study of such fields in spherical geometry in terms of what
is now known as Chandrasekhar--Kendall functions \cite{CK57}.
In the present paper we discuss the significance of these functions for
understanding the dynamics of magnetic fields exhibiting inverse
transfer behavior from small to large length scales.
This mechanism is significant not only in stars \cite{FPLM75,PFL76},
accretion discs \cite{BNST}, and galaxies \cite{Gre08},
but possibly even on the scale of the Universe in connection
with primordial magnetic fields \cite{BEO,CHB01}.

\section{Chandrasekhar--Kendall functions and application to Cartesian domains}

When it was realized that astrophysical magnetic fields might often be
force--free \cite{LS54}, i.e.\ $(\nab\times\BB)\times\BB=\bm{0}$,
Chandrasekhar began studying the mathematical properties of such fields
both in cylindrical and spherical coordinates \cite{Cha56_ff}.
Such fields are eigenfunctions of the curl operator, i.e.,
\EQ
\nab\times\hh_n=k_n\hh_n,
\EN
where the $k_n$ are eigenvalues and $\hh_n$ eigenfunctions.
These functions play important roles as basis functions for
solenoidal vector fields as well as for decomposing them
into right- and left-handed components.
This led to an important joint paper with Kendall \cite{CK57}, whose
supervisor (V C A Ferraro) remarked in a footnote of their paper that
``The results in this paper were derived independently by the two authors;
and they agreed to write it as one''.
Their work gained in importance since the 1970s, when force--free
magnetic fields were observed in the solar corona \cite{Low77} and produced
in laboratory plasmas \cite{Tay86}.

Chandrasekhar--Kendall or CK functions continue to be used in solar physics,
for example in describing the dynamics of coronal loops \cite{Vinod}.
These functions have subsequently been exploited also in other
geometries, including cylindrical geometries \cite{Mor07}, where they
are also referred to as Lundquist fields \cite{Linton1998}.
In the plasma physics community, these functions are  sometimes also
referred to as Chandrasekhar--Kendall--Furth or CKF functions \cite{CKF}.
In the following, however, we will focus on the Cartesian domains
with periodic boundaries \cite{Wal93}, where such functions are also
known as Beltrami waves.

In periodic Cartesian domains it is convenient to adopt CK functions
in Fourier space.
For example the magnetic vector potential can be written in Fourier
space (indicated by subscript $\kk$) as
\EQ
\AAA_{\kk}(t)=\int\AAA(\xx,t)\,e^{\ii\kk\cdot\xx}\,\dd^3\xx,
\EN
and it can be expanded in terms of eigenfunctions of the curl
operator with positive and negative eigenvalues, as well as a
longitudinal part that vanishes upon taking the curl to
compute the magnetic field $\BB=\nab\times\AAA$.
Thus, we write \cite{Wal93,CHB01,BDS02}
\EQ
\AAA_{\kk}(t)=a_{\kk}^+(t)\hh_{\kk}^++a_{\kk}^-(t)\hh_{\kk}^-
+a_{\kk}^\parallel(t)\hh_{\kk}^\parallel,
\EN
with
\EQ
\ii\kk\times\hh_{\kk}^{\pm} = \pm k \hh_{\kk}^{\pm},
\quad\quad k = |\kk|,
\EN
where $\hh_{\kk}^{\pm}$ are the eigenfunctions with positive and
negative eigenvalues, respectively.
These functions are orthonormal, i.e.,
\EQ
\bra{{\hh_{\kk}^+}^*\cdot\hh_{\kk}^-}
=\bra{{\hh_{\kk}^+}^*\cdot\hh_{\kk}^\parallel}
=\bra{{\hh_{\kk}^-}^*\cdot\hh_{\kk}^\parallel}=0
\EN
and
\EQ
\bra{{\hh_{\kk}^+}^*\cdot\hh_{\kk}^+}
=\bra{{\hh_{\kk}^-}^*\cdot\hh_{\kk}^-}
=\bra{{\hh_{\kk}^{\parallel}}^*\cdot{\hh_{\kk}^{\parallel}}}=1,
\EN
where asterisks denote the complex conjugate,
and angular brackets denote volume averages.
The longitudinal part $a_{\kk}^\parallel\hh_{\kk}^\parallel$ is
parallel to $\kk$ and vanishes after taking the curl
(or crossing with $\ii\kk$) to calculate the magnetic field, i.e., we have
\EQ
\BB_{\kk}=k\left(a_{\kk}^+\hh_{\kk}^+-a_{\kk}^-\hh_{\kk}^-\right).
\label{Bdecomp}
\EN
Magnetic helicity and magnetic energy spectra are then given by
\EQ
\AAA_{\kk}\cdot\BB_{\kk}^*=k\left[(a_{\kk}^+)^2-(a_{\kk}^-)^2\right],
\EN
\EQ
\BB_{\kk}\cdot\BB_{\kk}^*=k^2\left[(a_{\kk}^+)^2+(a_{\kk}^-)^2\right].
\EN
This implies evidently
\EQ
\BB_{\kk}\cdot\BB_{\kk}^*\ge k\left|\AAA_{\kk}\cdot\BB_{\kk}^*\right|,
\EN
which is also known as the realizability condition \cite{Mof69},
that is often quoted in terms of shell-integrated spectra of magnetic
energy and magnetic helicity, $M_k$ and $H_k$, respectively, i.e.,
\EQ
M_k\ge k|H_k|/2\mu_0.
\EN
Here, $\mu_0$ is the vacuum permeability and the $2\mu_0$ factor
comes from  the normalization of the magnetic energy spectrum,
$\int M_k\,\dd k=\bra{\BB^2/2\mu_0}$, while the magnetic helicity
spectrum is normalized such that $\int H_k\,\dd k=\bra{\AAA\cdot\BB}$.

In order to monitor the chiral dynamics of a system, it is convenient
to plot energy spectra of the contributions from right- and
left-handedly polarized parts.
Fortunately, this can be done without ever performing the explicit
decomposition \eq{Bdecomp} in terms of $a_{\kk}^\pm$.

\section{Application to the inverse cascade}

Eigenfunctions of the curl operator are well suited as
function basis for hydrodynamic turbulence \cite{Wal93}.
More recently they have also been used in magnetohydrodynamics
\cite{CHB01}, and in dynamo theory \cite{BDS02}, where
flows allow a conversion of kinetic into magnetic energy.

Dynamos are possible under completely isotropic conditions without helicity.
In that case we often speak about small-scale dynamos; see Ref.~\cite{BS05}
for a review.
However, in the presence of helicity, there is an inverse cascade of
magnetic helicity that leads to the gradual build-up of magnetic fields
at the largest scale of the system \cite{FPLM75,PFL76}.
This has been demonstrated in real space with turbulence driven
by an ABC-flow forcing function \cite{BP99} and, more clearly, with
turbulence driven by a forcing function proportional to
CK functions that are $\delta$-correlated in time \cite{B01}.
In the latter case, the inverse cascade process is clearly demonstrated
in terms of CK-decomposed spectra, $M_k^\pm$, as was done in Ref.~\cite{BDS02}.
They used a relatively big scale separation ratio, i.e., their forcing
wavenumber $\kf$ was large compared with the smallest wavenumber of the
domain, $k_1$ (they used $\kf/k_1=27$).
This implies, however, that the magnetic Reynolds number,
\EQ
\Rm=\urms/\eta\kf,
\EN
is smaller than if the scale separation was less.
It is therefore useful to increase the numerical resolution as
much as possible.
In this paper we present new simulations of helically driven isotropic
turbulence in a periodic domain at a resolution of $512^3$ meshpoints
and scale separation ratios of $\kf/k_1=15$ and 30.
The resulting magnetic Reynolds number reaches values up to $\Rm\approx57$
and the magnetic Prandtl number is unity.
The simulations have been performed with the
{\sc Pencil Code}\footnote{\texttt{http://pencil-code.googlecode.com}},
which is a modular high-order code (sixth order in space and third-order
in time, by default) for solving a large range of partial differential
equations.

\begin{figure}[t!]\begin{center}
\includegraphics[width=.49\columnwidth]{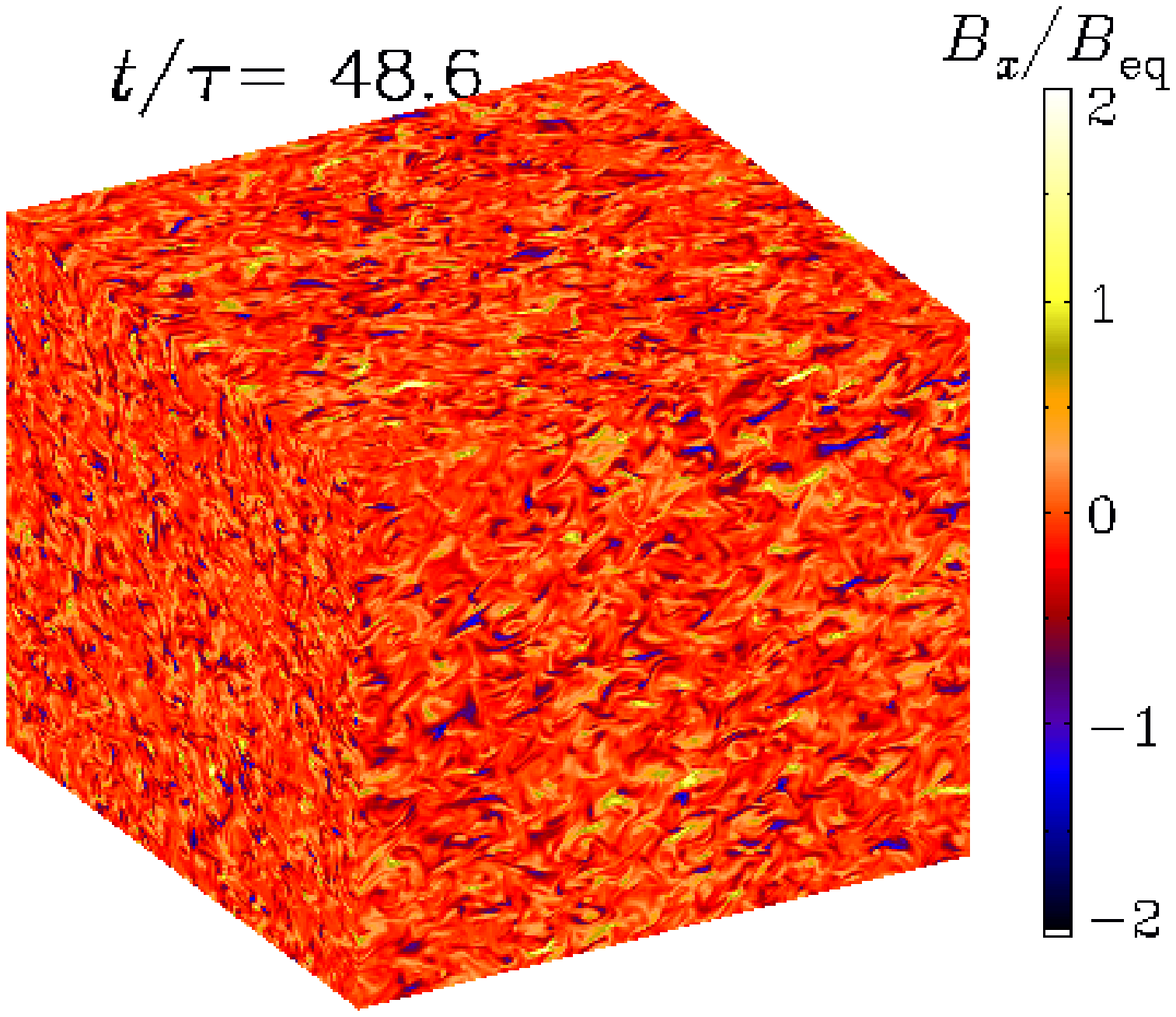}
\includegraphics[width=.49\columnwidth]{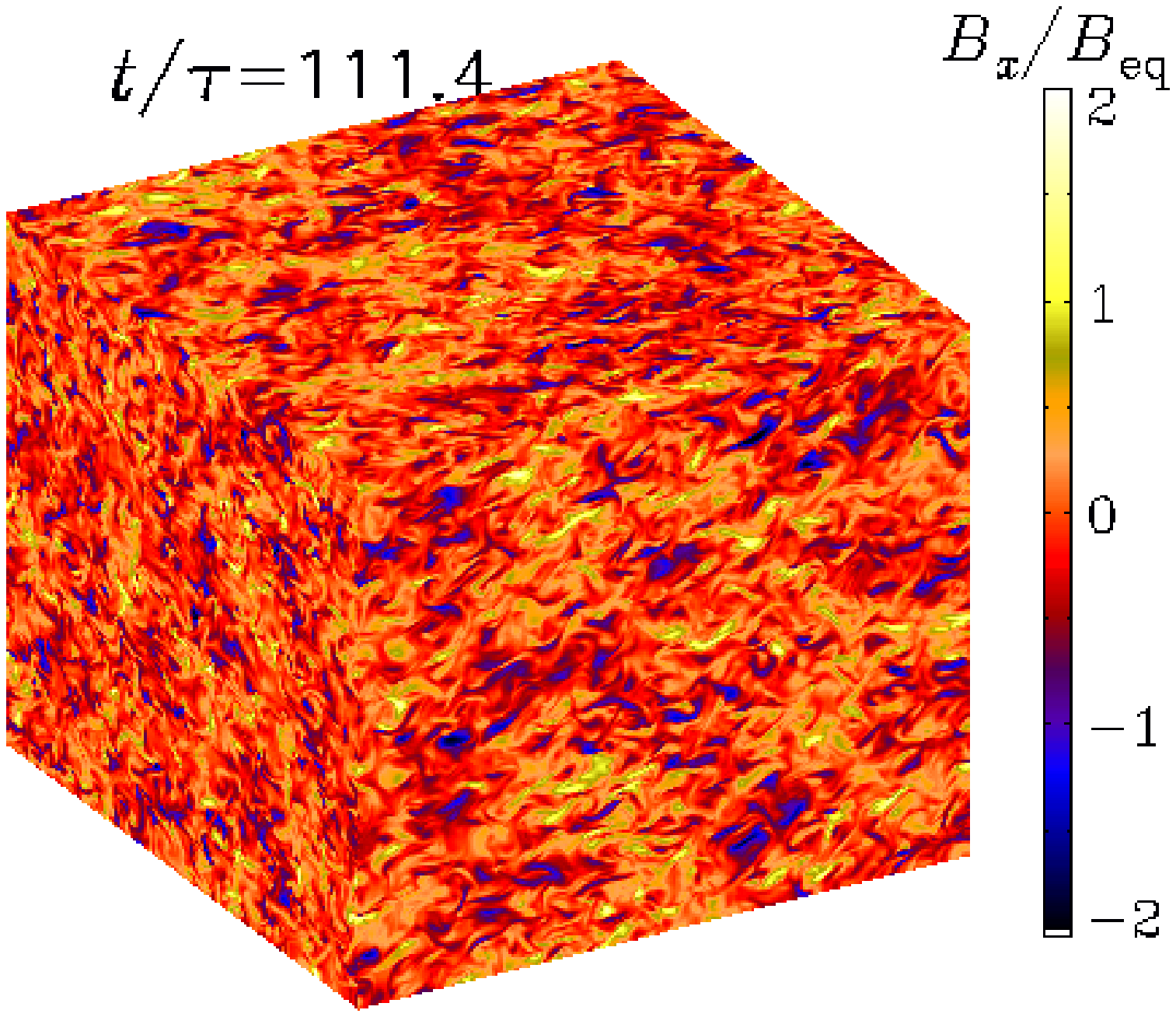}
\end{center}\caption[]{
Visualizations of $B_x/\Beq$ on the periphery of the domain at three
times when the small-scale magnetic field has already saturated.
Here, $B_{\rm eq}=\sqrt{\mu_0\rho_0}\,\urms$ is the equipartition
field strength where kinetic and magnetic energy densities are comparable,
and $\rho_0$ is the mean density.
Note that the maximum field strength is about twice $B_{\rm eq}$.
}\label{img1}\end{figure}

\begin{figure}[t!]\begin{center}
\includegraphics[width=\columnwidth]{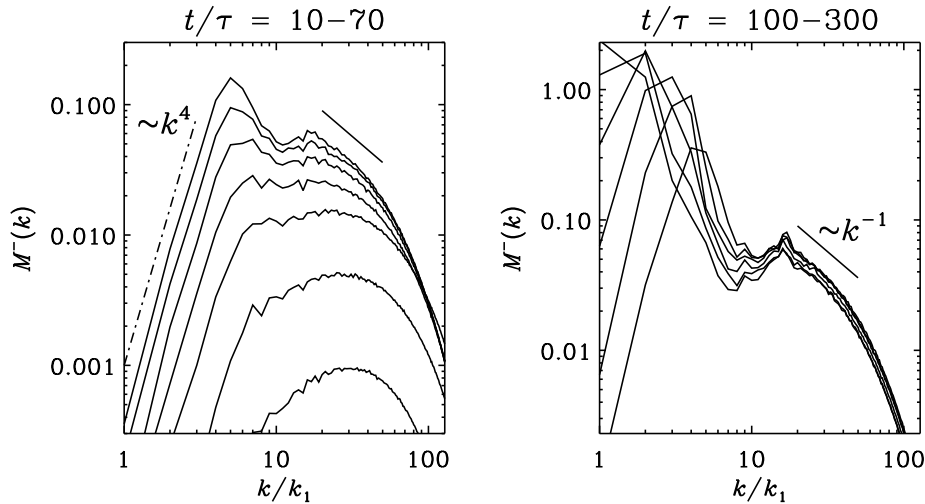}
\end{center}\caption[]{
Spectra of magnetic energy of the negatively polarized contributions,
$M^-_k$, at earlier (left) and later (right) times.
The scale separation ratio is $\kf/k_1=15$.
The range of time $t$ is given in  units of the turnover time,
$\tau=1/\urms\kf$.
At small wavenumbers, the $M^-_k$ spectrum is proportional to $k^4$,
while to the right of $\kf/k_1=15$ there is a short range with
a $k^{-1}$ spectrum.
}\label{pspec_ck}\end{figure}

In \Fig{img1} we show the $x$ component of the magnetic field at the
periphery of the domain at two early times in the simulation.
Note, that the magnetic field appears to organize itself gradually
into larger scale patches.
This is more clearly seen in energy spectra; see \Fig{pspec_ck}, where
we show the spectral magnetic energy at different times.
At early times the magnetic energy resides almost entirely at the
dissipative scale.
However, as time goes on, a magnetic field component at intermediate
scale with wavenumber $\kmean\approx\kf/2$ becomes discernible.
This corresponds to the fastest growing eigenmode of a  purely helical
large-scale dynamo \cite{BDS02}.
Note also that near the largest scale of the system, the magnetic
field has a power spectrum comparable with $k^4$, which corresponds
to white noise in the magnetic vector potential and obeys causality
\cite{DC03}.
To the right of the injection wavenumber,  $\kf$, there is a
short range of a $k^{-1}$ spectrum in $M^-_k$.

\begin{figure}[t!]\begin{center}
\includegraphics[width=.99\columnwidth]{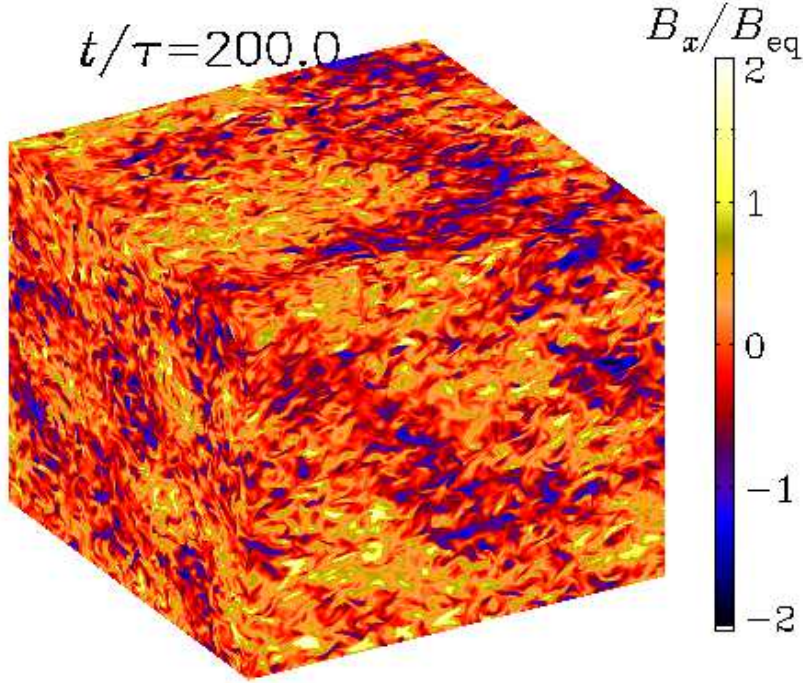}
\end{center}\caption[]{
Visualizations of $B_x/\Beq$ on the periphery of the domain at a later
time when a large-scale magnetic field is beginning to emerge.
This time corresponds to the intermediate time shown in the second
panel of \Fig{pspec_ck} where
the peak of the spectrum has reached the wavenumber $\kmean/k_1\approx3$,
corresponding to about 3 large-scale patches across the domain.
The scale separation ratio is $\kf/k_1=15$, so there are about 15 eddies
across the domain.
}\label{img2}\end{figure}

\begin{figure}[t!]\begin{center}
\includegraphics[width=\columnwidth]{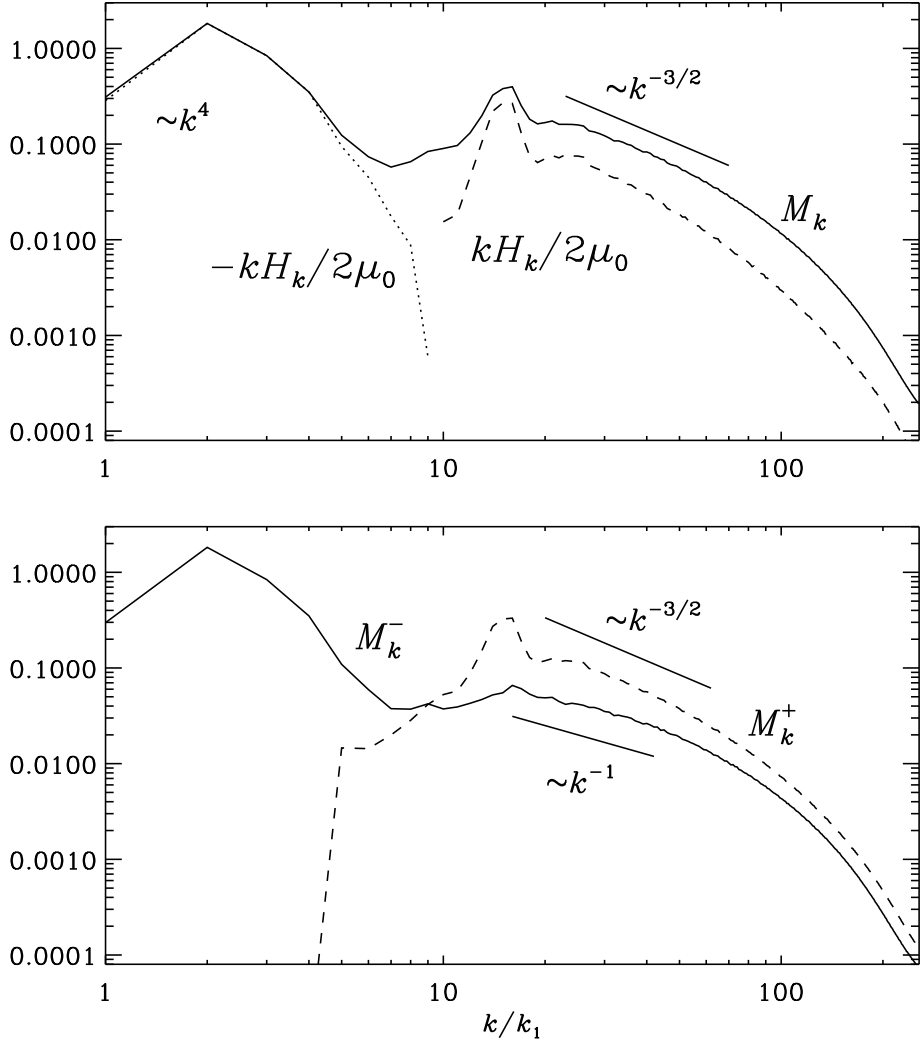}
\end{center}\caption[]{
Upper panel: spectra of magnetic energy, $M_k$,
and rescaled magnetic helicity, $\pm kH_k/2\mu_0$.
Lower panel: spectra of magnetic energy of positively and negatively
polarized parts, $M_k^\pm$.
}\label{pspec_ck0}\end{figure}

At later times, magnetic energy is transported further to smaller
wavenumbers; see the right-hand panel of \Fig{pspec_ck},
where the peak of the spectrum at $k\equiv\kmean$ travels to
smaller wavenumbers.
The effect of this is also evident in real space in that larger
scale patches of magnetic field with the same orientation occur.
In \Fig{img2} we show the $x$ component of the magnetic field at the
time when a large-scale magnetic field has emerged, but it has still
not reached the scale of the computational domain.
Magnetic energy spectra of that time are shown in the upper panel
of \Fig{pspec_ck0}.
Note, however, that the magnetic energy does show a clearly pronounced
secondary peak that is now at wavenumber $\kmean/k_1\approx2$.

\begin{figure}[t!]\begin{center}
\includegraphics[width=\columnwidth]{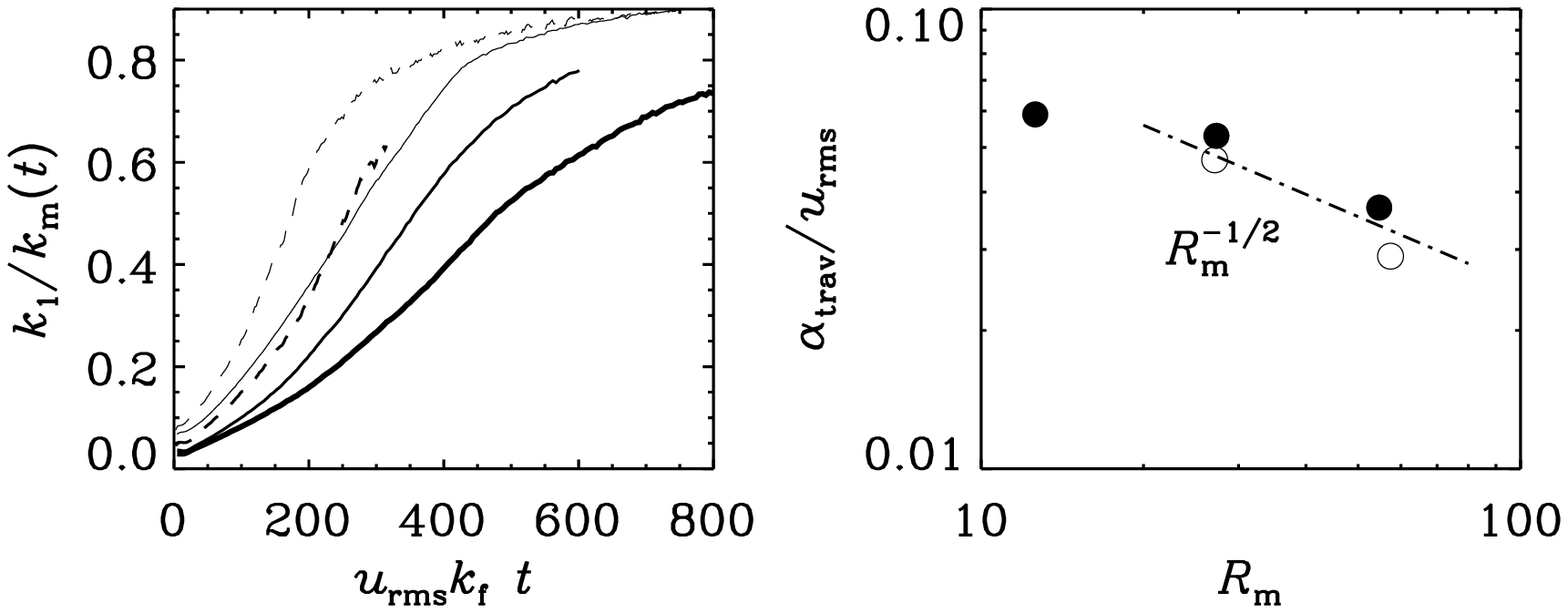}
\end{center}\caption[]{
Left panel: time dependence of the peak wavenumber for scale separation
ratios of 15 (dashed) and 30 (solid lines) at $\Rm$ of 12, 27, and
around 57 (increasing line thickness).
Right panel: $\Rm$ dependence of the cascade speed for scale separation
ratios of 15 (open symbols) and 30 filled symbols.
The straight lines give the $\Rm^{-1/2}$ (dotted) and
$\Rm^{-1}$ (dashed) dependences.
}\label{palp_trav}\end{figure}

The process that produces magnetic fields at large length scales
shown in the present paper can loosely be regarded as an inverse cascade of
magnetic helicity and thereby magnetic field.
However, it can also be treated in terms of mean-field theory, in which
case it corresponds to the so-called $\alpha$ effect, whose physics
was already identified by Parker \cite{Par55}.
This dynamo is quite different from the laminar dynamos that were
studied at the time by Bullard and Gellman \cite{BG54}, although
there remained substantial uncertainty about the convergence of
the numerical solutions obtained back then.

One of the surprising results of the last 10 years is the fact that
in a closed or periodic domain, the large-scale magnetic field saturates
only on a resistive time scale \cite{B01}.
This is understood as a consequence of the magnetic helicity equation,
which limits the time scale for magnetic helicity evolution to the
resistive time \cite{Ber84}, and lowers the nonlinear $\alpha$ effect
in an $\Rm$--dependent fashion \cite{KR82,FB02,BB02,Sub02}.
According to this picture, if we associate, as was done in Ref.~\cite{B01},
the peak wavenumber $\kmean$
of the inversely cascading field with the $\alpha$ effect via
$\kmean=\alpha/2\etat$, where $\etat$ is the turbulent magnetic diffusivity,
we should expect the speed of inverse transfer to slow down as well.
When this was first proposed \cite{BDS02}, the results were not
fully conclusive and of course in conflict with the usual understanding
that the inverse cascade should operate on a dynamical time scale
\cite{Ver01}.
Our present results allow us now a fresh look at this problem.
As a continues measure of $\kmean$ we define
\EQ
\kmean^{-n}=\int k^{-n}\,M_k\,\dd k\left/\int M_k\,\dd k\right.,
\EN
where we choose $n=3$, which was found to give the best approximation
to the actual location of the secondary peak in $M_k$.
This is probably related to the sharp rise of $M_k$ proportional to $k^4$
for $k_1<k<\kmean$.
Following earlier work \cite{BDS02}, the evolution of $\kmean(t)$ is
well described by the fit formula
\EQ
\kmean^{-1}=\alpha_{\rm trav}(t-t_{\rm sat}),
\EN
where the parameter $\alpha_{\rm trav}$ characterizes the speed at which
$\kmean$ travels toward smaller wavenumbers.
This parameter is not  directly related to the $\alpha$ effect,
although $\alpha_{\rm trav}$ also has the dimension of velocity.
The result is shown in \Fig{palp_trav}, where we plot $\kmean^{-1}$
versus time for different values of $\Rm$ and $\kf/k_1$, and we
also show the dependence of the slope $\alpha_{\rm trav}$ on $\Rm$.
In agreement with earlier work \cite{BDS02}, $\alpha_{\rm trav}$
is clearly $\Rm$ dependent and decreases with increasing $\Rm$
roughly proportional to $\Rm^{-1/2}$.
The reason for this seems clear: the bump in the spectrum
is a manifestation of an $\alpha$ effect causing {\it nonlocal}
spectral transfer to smaller $k$, but it is not an inverse cascade
in the usual sense.
An inverse cascade may still occur at intermediate wavenumbers
$\kmean\ll k\ll\kf$, where evidence for constant ($k$-independent)
spectral magnetic helicity transfer to larger scale has been
found \cite{MM10}.

\section{Closing the loop: turbulence from the magnetic field itself}

Large-scale dynamos show mean magnetic fields that can either be
statistically steady, or they can oscillate and propagate in space.
In the language of mean-field dynamos, the former ones are typically
the so-called $\alpha^2$ dynamos while the latter ones are the
$\alpha\Omega$ dynamos, where the $\Omega$ effect corresponds
to differential rotation or shear amplifying the toroidal magnetic
field from a poloidal one.
A prominent example of such a dynamo is that believed to exist in
accretion discs, where shear plays a strong role \cite{BNST}.
This brings us to our final example where Chandrasekhar's contributions
are still being mentioned in modern astrophysics, namely in connection
with the magneto-rotational instability.

Already in 1953, Chandrasekhar studied the hydromagnetic stability
of Couette flows, e.g.\ the flow of a liquid metal between two rotating
cylinders with angular velocities $\Omega_1$ and $\Omega_2$ at radii
$R_1$ and $R_2>R_1$.
In view of astrophysical applications to the stability of accretion
discs, particularly interesting is the case where angular velocity
decreases with radius, i.e.\ $\Omega_1>\Omega_2$, but angular momentum
increases, i.e.\ $R_1^2\Omega_1<R_2^2\Omega_2$.
Under the assumption of a small magnetic Prandtl number,
where the viscosity is much smaller than the magnetic diffusivity,
and a small gap width, Chandrasekhar \cite{Cha53} found that the flow is
stable both with and without an axial magnetic field.
However, relaxing the assumption of a small gap width, Velikhov \cite{Vel59}
found that for an ideal fluid (zero viscosity and zero magnetic diffusivity)
the flow is unstable when $\Omega_1>\Omega_2$.
This was the also confirmed by Chandrasekhar \cite{Cha60}, but the
astrophysical significance was hardly appreciated until Balbus \& Hawley
\cite{BH91} rediscovered this instability, which is now usually called
the magneto-rotational instability or MRI.

The point is that both Velikhov and Chandrasekhar were very much
ahead of their time, because accretion discs entered the astrophysical
literature only in the late 1960s and early 1970s.
Nevertheless, it is important to realize that in the meantime the
MRI was not completely forgotten.
In his book, Safronov \cite{Saf72} discussed several mechanisms for causing
turbulence in {\it protostellar} discs.
Among the various mechanisms, he did discuss the MRI, but discarded it
on the grounds that protostellar discs are cold and poorly ionized,
so magnetic fields would not be tied to the gas.
His work was before the famous paper of Shakura \& Sunyaev \cite{SS73},
where the physics of accretion discs was applied to the much
hotter discs around stellar-mass and supermassive black holes.
Curiously enough, until then the focus has rather been in
stability, as is evidenced by a quote from S H Hong \cite{SHHong76}
who writes ``A special form  of this problem was first investigated by
Chandrasekhar \cite{Cha53}, and later by Velikhov in \cite{Vel59}.
On the basis of the
small gap approximation, they found that the effect of a sufficiently
strong magnetic field is to inhibit the onset of instability.''

The MRI is now generally believed to be the main agent that drives
turbulence in accretion discs, and this turbulence is able to reinforce
the magnetic field by dynamo action via a doubly-positive feedback
\cite{BNST,Stone,ZR01,Gre10}.
These discoveries would not have been possible without the help of
supercomputers.
However, coming back to Chandrasekhar, it is fair to say that he always
had an eye toward numerical solutions.
For example his approach to the dynamo problem in 1956, where he found
a significant slow-down of the decay \cite{Cha56_slowdecay}, was also
entirely based on numerical solutions.
It is therefore clear that his combined approach using numerical and
analytical tools was not only a modern one, but much of his work attracts
interest still today, as is evidenced by the increasing citation rate
of his book on {\it Hydrodynamic and hydromagnetic stability} \cite{Cha61}.
\\

I acknowledge the allocation of computing resources provided by the
Swedish National Allocations Committee. This work was supported in part by
the European Research Council under the AstroDyn Research
Project No.\ 227952 and the Swedish Research Council Grant No.\ 621-2007-4064,
and the National Science Foundation under Grant No.\ NSF PHY05-51164.

\bibliographystyle{pramana}
\bibliography{references}

\end{document}